\documentclass[runningheads,a4paper]{llncs}

\usepackage{amssymb}
\usepackage[table]{xcolor}
\usepackage{graphicx}
\usepackage{algorithm}
\usepackage{algorithmic}
\usepackage{multirow}
\usepackage{makecell}
\usepackage{url}

\urldef{\mailsa}\path|{kejriwal,miranker}@cs.utexas.edu|    
\newcommand{\keywords}[1]{\par\addvspace\baselineskip
\noindent\keywordname\enspace\ignorespaces#1}

\begin{document}

\mainmatter  

\title{Self-contained NoSQL Resources for Cross-Domain RDF}

%
%
\author{Mayank Kejriwal%
\and Daniel P. Miranker}

\institute{University of Texas at Austin\\
\mailsa}

%
%

\toctitle{Lecture Notes in Computer Science}
\tocauthor{Authors' Instructions}
\maketitle

\begin{abstract}
Cross-domain knowledge bases such as DBpedia, Freebase and YAGO have emerged as encyclopedic hubs in the Web of Linked Data. Despite enabling several practical applications in the Semantic Web, the large-scale, schema-free nature of such graphs often precludes research groups from employing them widely as evaluation test cases for entity resolution and instance-based ontology alignment applications. Although the ground-truth linkages between the three knowledge bases above are available, they are not amenable to resource-limited applications. One reason is that the ground-truth files are not self-contained, meaning that a researcher must usually perform a series of expensive joins (typically in MapReduce) to obtain usable information sets.
In this paper, we upload several publicly licensed data resources to the public cloud and use simple Hadoop clusters to compile, and make accessible, three cross-domain self-contained test cases involving linked instances from DBpedia, Freebase and YAGO. Self-containment is enabled by virtue of a simple NoSQL JSON-like serialization format. Potential applications for these resources, particularly related to testing transfer learning research hypotheses, are also briefly described.   
     
\end{abstract}

\section{Introduction}\label{introduction}
Cross-domain knowledge bases have emerged as \emph{structured encyclopedias} in the Linked Open Data ecosystem \cite{comparative}. Examples include Freebase, DBpedia, and YAGO \cite{freebase}, \cite{dbpedia}, \cite{yago}, but also industry-driven efforts such as projects by Microsoft (e.g. Satori\footnote{\url{http://www.bing.com/blogs/site_blogs/b/search/
archive/2013/03/21/satorii.aspx}}), Facebook\footnote{\url{https://en.wikipedia.org/wiki/Facebook_Graph_Search}} and Google (e.g. the Google Knowledge Graph\footnote{\url{http://www.google.com/insidesearch/features/
search/knowledge.html}}). Such knowledge bases are described as \emph{cross-domain} as they contain a large set of types that cannot be classified under a single domain, but are best visualized as a collection of interlinked domains \cite{comparative}. As such, they are intended to provide a broad coverage of ontological types, properties, instances and URIs.  

The broad vocabularies of these cross-domain graphs make them useful in a variety of Semantic Web applications. DBpedia, for example, is among the most densely connected RDF datasets on Linked Open Data, and serves as a `hub' connecting many myriad domains \cite{dbpedia}. Linking a new RDF graph to DBpedia is a vital starting point in fulfilling the fourth Linked Data principle, which states that data should be interlinked rather than published in silos \cite{linkeddata}. Other applications include using these knowledge bases to build high-fidelity \emph{knowledge vaults}, which can be used to support large-scale semantic search \cite{kvault}, and also \emph{knowledge graph completion} \cite{kgi}. 

Many of these applications require solutions to problems such as entity resolution (ER), which is the algorithmic problem of finding and connecting pairs of entities (in one or more knowledge bases) that refer to the same underlying entity \cite{rahmsurvey}. State-of-the-art ER systems tend to be mostly serial, and evaluations are often limited to single-domain datasets \cite{datamatching}, \cite{rahmsurvey}. Developing ER solutions for large-scale, heterogeneous data has already been recognized as an important problem, but evaluations involving multiple cross-domain graphs are rare \cite{hhis}. 

One possible bottleneck is \emph{data preparation}. In the case of knowledge bases like DBpedia and YAGO, the instances, types and ground-truths are usually described by multiple files (Section \ref{resources}), which can together be tens of gigabytes in size \cite{dbpedia}, \cite{yago}. Data preparation involves linking these files together to derive the information sets required by an ER system. Such linking involves NoSQL (often, MapReduce-based) joins that are error-prone in practice, and involve significant effort. In the case of knowledge bases like Freebase, scale can be an even bigger issue \cite{freebase}. The uncompressed Freebase dump, which is a single N-triples file, is hundreds of gigabytes in size\footnote{\url{https://developers.google.com/freebase/data#freebase-rdf-dumps}}, not including extraneous information such as deleted triples. There is significant cost and effort involved in uploading such a dataset to a cluster, and designing MapReduce algorithms that can accommodate \emph{data skew} and \emph{curse of last reducer} issues \cite{curse}. 

Together, these observations motivate the development of cross-domain evaluation and ground-truth files that are (1) relatively \emph{self-contained}, that is, do not involve joins, and (2) are easy to access and use, and can support a broad research agenda. In this paper, we present three resource files in support of these requirements. A line in each resource describes a (2-way or 3-way) link, and can be parsed by simple non-recursive code independent of other lines. Non-recursive self-containment is achieved by encapsulating each entity participating in the link in a flat NoSQL JSON-like data structure that is further described, along with the Hadoop-based resource generation pipeline, in Section \ref{resources}. Section \ref{applications} concludes the work by listing two research questions, directly connected to these resources, that we are currently investigating. 

{\bf Related Work.} Knowledge bases are fundamental to the Semantic Web vision \cite{semanticweb}. The Web of Linked Data has emerged as a particular success story, and when last surveyed, contained many billions of triples in domains ranging from social media to cross-domain \cite{linkeddata}. In the Linked Open Data diagram\footnote{\url{http://linkeddata.org/}}, cross-domain datasets such as DBpedia and Freebase are at (or close to) the center on account of their high connectivity to other knowledge bases. F{\"a}rber et al. compare and describe popular knowledge bases in a recent survey \cite{comparative}.

Because of its importance to many applications, \emph{entity resolution} (ER) was chosen as the primary task for which the resources in this paper were developed. A good survey was provided by K{\"o}pcke and Rahm \cite{rahmsurvey}. There has been a recent trend in the ER community, particularly in the Semantic Web, to present novel benchmarks for ER evaluations \cite{benchmarks}. This work is in support of that trend. All datasets and resources developed in this paper are real-world and public.

\section{Self-Contained NoSQL Ground-truths}\label{resources}
NoSQL data structures have become increasingly popular in recent years, especially in the Semantic Web \cite{nosql}. The importance of XML was recognized even in the early days \cite{semanticxml}. More recently, JSON-LD\footnote{\url{http://json-ld.org/}} was proposed as a lightweight Linked Data format. The basic goal of JSON-LD is to provide a way for JSON data to interoperate at the Web scale. 

\begin{figure*}[t]
\centering
\includegraphics[height=5.6cm, width=14.0cm]{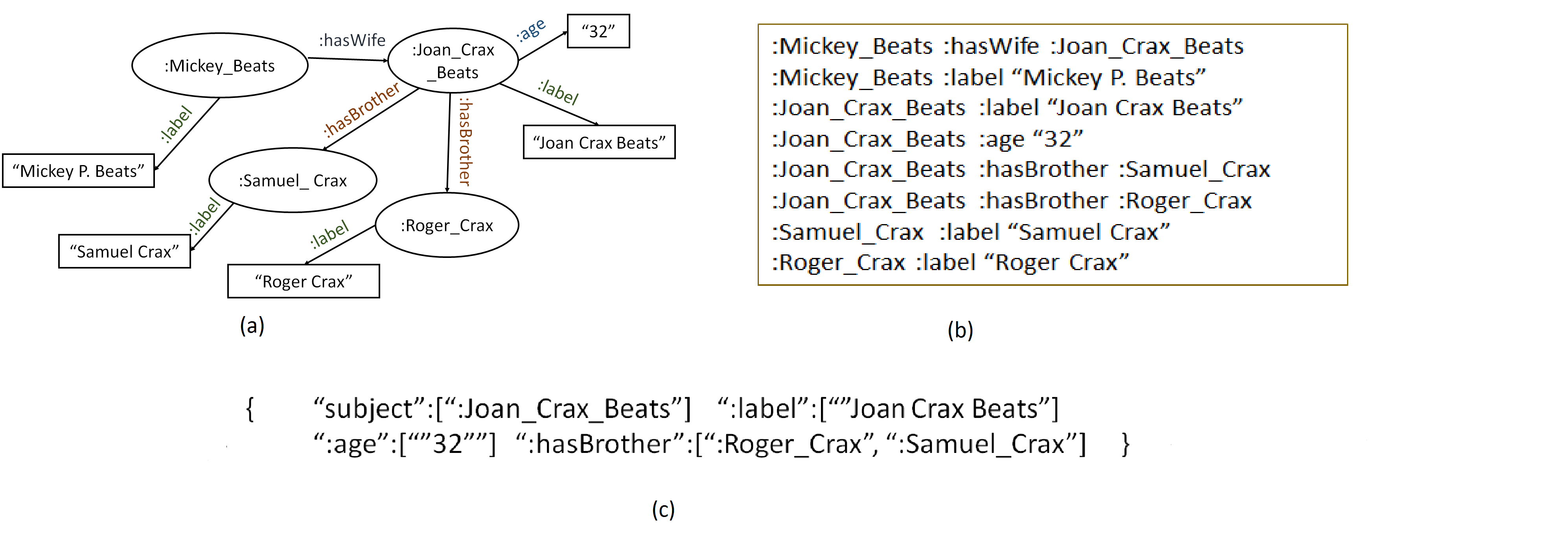}
\caption{An RDF dataset equivalently represented as a visualized, directed graph (a) and as a set of triples (b). (c) encodes the information set of the entity \emph{:Joan\_Crax\_Beats} as a flat JSON-like data structure with tab-delimited key-value pairs (see text for syntax) }
\label{nosql-example} 
\end{figure*}
For the resources in this paper, we designed a similar lightweight format that is JSON-like and contains the same information set as a \emph{logical property table} representation of RDF that we had earlier proposed\footnote{We provide formal definitions of this data structure in \cite{kejriwaljournal}.} \cite{lpt}, inspired closely by the physical data structure implemented in triple stores such as Jena \cite{propertytable}. Figure \ref{nosql-example} provides an illustration of the JSON-like data structure proposed herein. Each entity is represented as a set of \emph{key-value} pairs, where the key is always a string and a value is always a list of strings\footnote{Syntactically, both URI and literal object values are treated as strings. We distinguish the latter from the former by enclosing literals within two double quotes (e.g. ````32"" in Figure \ref{nosql-example}).}. We use a list to hold the values, because a property can often have more than one object value (e.g. in Figure \ref{nosql-example} Joan Crax has two brothers).

Most importantly, the structure is \emph{flat}, and does not require a recursive parsing algorithm. Individual key-value pairs are \emph{tab-delimited}, and a single line of code in a high-level language such as Java can be used to parse the structure into an array of key-value pairs. We note that the vast majority of ER systems only consider entity \emph{labels} and \emph{surrounding} objects values as useful information describing a given entity\footnote{In a machine learning framework, features would be extracted from two such information sets (representing two entities), and a classifier would be used to score the entity pair as being linked or not linked \cite{SWcomparison}.} \cite{rahmsurvey}; the proposed NoSQL data structure captures all of this information in a single self-contained line. 

Given this simple serialization, the contributed resources were constructed as follows. First, we accessed and downloaded a subset of public raw files describing the English DBpedia and YAGO, and the entire Freebase dump (downloaded in mid-2015; see footnote 4 for the link). While the Freebase dump was a single N-Triples file (not unlike Figure \ref{nosql-example} (b)) on the order of hundreds of gigabytes, DBpedia and YAGO comprised multiple files. Specifically, for DBpedia we downloaded the \emph{infobox\_properties} and \emph{instance-types} files (in both cases, English only)\footnote{A list of all available files, versions and languages may be browsed at  \url{http://wiki.dbpedia.org/Downloads2015-10}}. For YAGO, the available files were more fine-grained\footnote{\url{http://www.mpi-inf.mpg.de/departments/databases-and-information-systems/research/yago-naga/yago/downloads/}}, and we downloaded separate files describing instances, labels, literal facts, data facts and (non-literal and non-date) facts. Ground-truth files connecting entities between each of these three knowledge bases are already available\footnote{A description of the methodology describing how these original ground-truths were generated to begin with, may be found at \url{http://wiki.freebase.com/wiki/DBPedia}. The key technique is exploiting Wikipedia as a mutual source from which many facts in these knowledge bases were extracted.} and may also be accessed at those links. Note that the ground-truths are not \emph{self-contained}, and typically only specify pairs of URIs (formally represented as \url{:sameAs} triples), which motivated the developments in this work. 

\begin{figure*}[t]
\centering
\includegraphics[height=4.3cm, width=12.0cm]{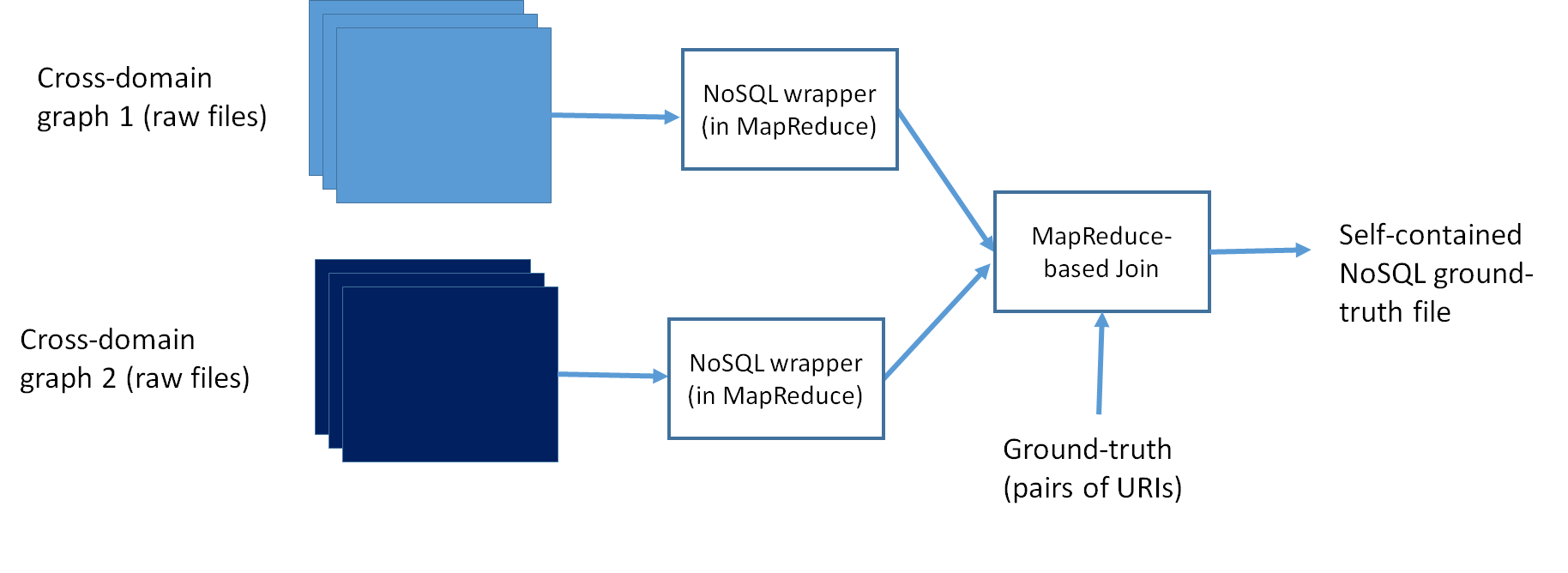}
\caption{A high-level illustration of the Hadoop pipeline used to construct the resource}
\label{pipeline} 
\end{figure*}
The raw files were uploaded to public cloud storage (Microsoft Azure) and a Hadoop pipeline was built to process these files. Figure \ref{pipeline} illustrates the process at a high-level. First, the raw files (for each cross-domain graph) are compiled by a `wrapper' program into a single file, where each line describes an entity\footnote{From the perspective of the wrapper programs, a URI is an entity iff it occurs as a subject in at least one triple.} using the previously described flat NoSQL data structure. The two files thus output by the wrappers are input to a second MapReduce job, which takes as additional input the (previously available) ground-truth file that contains pairs of URIs that represent positive links. A three-way \emph{inner join} is then performed on the three files. The final result is a self-contained ground-truth file, the format of which is described below.

We executed the pipeline \emph{thrice} using Microsoft Azure HDInsight clusters (ranging in size from 6-14 A3 nodes\footnote{\url{https://azure.microsoft.com/en-us/pricing/details/hdinsight/}}) and obtained three ground-truth files, denoted henceforth as Freebase-DBpedia, DBpedia-Freebase-YAGO and YAGO-DBpedia. Freebase-DBpedia is a 12.98 GB file that contains 2,093,007 lines, with each line having the following format:\\
\emph{$<$Link ID$>$ freebase-instance $<$freebase-instance$>$ dbpedia-instance $<$dbpedia-instance$>$}

Namely, each line contains five tab-delimited `slots'. The first slot is for a system-generated link ID that is unique for each pair of URIs in the ground-truth. This is followed by the string `freebase-instance', which is followed by the (NoSQL-represented) Freebase instance participating in the link. This, in turn, is followed by the string `dbpedia-instance' and the corresponding DBpedia instance to which the Freebase instance is linked, per the ground-truth.

YAGO-DBpedia is of size 5.29 GB and also has format similar to Freebase-DBpedia. In total, YAGO-DBpedia has 2,554,643 lines. The size difference between Freebase-DBpedia and YAGO-DBpedia shows that YAGO instances are much more \emph{compact} compared to Freebase instances, which we observed to contain many long, opaque URIs.

Finally, DBpedia-Freebase-YAGO is of size 11.54 GB and contains 1,655,565 lines, with each line having a format that is an `extended' version of Freebase-DBpedia. Each line now contains a 3-way link, with one instance each from DBpedia, Freebase and YAGO (in that order). Note that it is possible for an instance (from any of these datasets) to participate in more than one link. We built this file by performing an additional join on the other two resource files (hence, the number of lines is strictly less than the minimum of the two). Rather than generate a unique ID for each 3-way link, we expressed the ID as a pair of IDs (from each of the two resource files), so that the provenance of each line is traceable, and the resource files can be used in tandem\footnote{Especially important when testing \emph{transfer learning} hypotheses (see Section \ref{applications}).}. 

All three resources have been uploaded to the Figshare public repository and may be publicly accessed and downloaded\footnote{The access link is \url{https://figshare.com/articles/Self_contained_ground_truths_for_cross_domain_linkage/3204325}}. Due to size limitations, we compressed the three resource files into a 4.68 GB zip file. As a best practice guideline, we have also included (in the resource space, but not in the zip) a 64 MB file that contains 10,000 randomly sampled lines from the Freebase-DBpedia resource. This is meant to aid serial experimentation, as well as benefit users who wish to download and experiment with a smaller version of the resource. 

We have also included documentation on the format of these files, earlier described, as well as syntax rules and parsing instructions for the flat NoSQL data structure. The resource also has a DOI citation\footnote{\url{https://dx.doi.org/10.6084/m9.figshare.3204325.v1}}. We plan to actively maintain and update the resource by periodically re-executing the pipeline in Figure \ref{pipeline}. We also hope to develop new wrappers, as existing cross-domain knowledge bases are updated, and new ones are released. 
 
Finally, note that the resources have been released under a CC-BY license. Because these files were ultimately derived from raw files and previously compiled ground-truths part of the Freebase, DBpedia and YAGO projects, their licensing requirements may apply, depending on the terms and purpose of use. These requirements may be especially applicable to organizations wishing to use the resources in a commercial setting.

\section{Applications}\label{applications}

The resources can be used to test several research hypotheses, some of which we are already investigating:

{\bf Transfer-learning across domains:} A cursory analysis of the resource files shows that the links span a wide range of domains, from football players to civil parishes. Both feature extraction and labeled training data are known to be important prerequisites in supervised machine learning applications \cite{rahmsurvey}. Given links across two domains, an interesting question is if features and training data can be transferred from a source domain to a target domain to improve performance on the latter. This is a tricky issue, as transfer learning can sometimes degrade performance. The only work on transfer learning (in the Semantic Web) for entity resolution that we are aware of concerns same-domain transfer \cite{machine1}.

{\bf Transfer-learning across datasets:} One of the resource files we presented contains 3-way links between DBpedia, Freebase and YAGO. This leads to the question of whether we can train a classifier on links between, say, DBpedia and Freebase, and then use the classifier to label links between Freebase and YAGO. To enable such an application, we would have to locate a common set of features that works well across both pairs of datasets. We would also need a uniform feature representation. 

Despite these challenges, we believe that transfer learning holds tremendous promise for Linked Open Data, which exhibits much variety.
We note that such hypotheses do not necessarily have to be explored in MapReduce, even though the resource files are several gigabytes in size. For some of our transfer learning experiments, we sampled `development' sets from the resource files using serial sampling algorithms. For example, we used the \url{rdf:type} property to filter instances that belonged to a specific domain (e.g. football player). By filtering instances of various types, we have a set of (different-domain) test cases that can be used to evaluate transfer learning serially. Such sampling procedures are only possible because each line in the resource file is self-contained. Algorithmically, each line can be read into main memory and processed on an individual basis.  

Beyond entity resolution, we also used these resources in a recent case-study involving large-scale ontology alignment between DBpedia and Freebase (currently under preparation). A good reference for this task, using locality sensitive hashing (LSH) techniques and also involving DBpedia and Freebase, is the work by  \cite{ibmlsh}. This effort was largely based in industry (IBM research); as noted in the introductory section, the use of large-scale cross-domain graphs in research has been rare, owing to bottlenecks of scale and data preparation. The hope is that the resources described in this work will help in addressing this issue. 

We believe that the resources can also find additional uses in industry, particularly as a structured semantic lexicon for knowledge base completion tasks. A domain-specific example (concerning population and normalization of knowledge bases describing \emph{companies}) of such a use may be found in \cite{company}. With these resources, we hope that more such applications can be enabled, especially by smaller, resource-strapped organizations looking to adopt semantic technology.

\section*{Acknowledgements}
The authors were supported by a US National Science Foundation grant.

\bibliography{typeinst}
\bibliographystyle{abbrv}
\end{document}